\documentclass[twocolumn, amsmath,amssymb,aps,floatfix,nofootinbib]{revtex4}

\usepackage{graphicx}
\usepackage{xcolor}
\usepackage{dcolumn}
\usepackage{bm}
\usepackage{siunitx} 
\usepackage{hyperref}
\usepackage{comment}
\usepackage{amssymb}
\usepackage[normalem]{ulem}

\usepackage{url}
\usepackage{amsmath,amssymb,graphicx}

\usepackage{lipsum} 

\usepackage[switch]{lineno}

\usepackage[utf8]{inputenc}
\usepackage{amsfonts}
\usepackage{multirow}
\usepackage{booktabs}
\usepackage{nicefrac}

\usepackage{xcolor}
\usepackage[export]{adjustbox}
\usepackage{mdframed}
\usepackage{ bbold }

\newcommand{\red}[1]{#1}

\begin{document}

\title{Super-Resolution Time-Resolved Imaging using Computational Sensor Fusion}

\author{C. Callenberg$^{1}$, A. Lyons$^{2}$,  D. den Brok$^1$, A. Fatima$^{2}$, A. Turpin$^{3}$, \red{V. Zickus$^{2}$},  L. Machesky$^4$, J. Whitelaw$^4$, D. Faccio$^2$, M.B. Hullin$^1$}

\affiliation{$^1$Institute of Computer Science, University of Bonn, Germany}
\affiliation{$^2$School of Physics \& Astronomy, University of Glasgow,  G12 8QQ Glasgow, United Kingdom}
\affiliation{$^3$School of Computing Science, University of Glasgow, G12 8LT Glasgow, United Kingdom}
\affiliation{$^4$Cancer Research UK, Beatson Institute, Glasgow, United Kingdom}
\affiliation{Corresponding author: ashley.lyons@glasgow.ac.uk / hullin@cs.uni-bonn.de / daniele.faccio@glasgow.uk}

\begin{abstract}
Imaging across both the full transverse spatial and temporal dimensions of a scene with high precision in all three coordinates is key to applications ranging from LIDAR to fluorescence lifetime imaging. However, compromises that sacrifice, for example, spatial resolution at the expense of temporal resolution are often required, in particular when the full 3-dimensional data cube is required in short acquisition times. We introduce a sensor fusion approach that combines data having low-spatial resolution but high temporal precision gathered with a single-photon-avalanche-diode (SPAD) array with set of data that has high spatial but no temporal resolution, such as that acquired with a standard CMOS camera. Our method, based on blurring the image on the SPAD array and computational sensor fusion, reconstructs time-resolved images at significantly higher spatial resolution than the SPAD input, upsampling numerical data by a factor 12$\times$12, and demonstrating up to 4$\times$4 upsampling of experimental data. We demonstrate the technique for both LIDAR applications and FLIM of fluorescent cancer cells. This technique  paves the way to high spatial resolution SPAD imaging or, equivalently, FLIM imaging with conventional microscopes at frame rates accelerated by more than an order of magnitude.
\end{abstract}

\maketitle

{C}{onventional} cameras produce images that show static illumination in a depicted scene, because exposure times are usually much longer than the photon transit time. Time-of-Flight (ToF) imaging systems, however, reach temporal resolutions of picoseconds or less and can therefore record the propagation of light in the scene. Obtaining both a high temporal and spatial resolution is particularly important for ToF imaging systems, which are often limited by the resolution in the time domain. A simple example are LIDAR-based systems where the measurement resolution in the temporal domain directly equates to the spatial depth resolution. In more complex applications, such as non-line-of-sight imaging (NLOS), the resolution of the ToF of the photons is critical for determining an object's position in all three spatial dimensions \cite{gariepy2015single,Velten2012,Buttafava2015,Faccio2018}. Access to high-resolution temporal information is also pertinent to challenges such as imaging through complex media \cite{Durduran2010,Satat2016,Lyons2019b}, and fluorescence lifetime imaging (FLIM), where the fluorophores of a target object are identified from their fluorescence lifetimes \cite{Dowling1999}.

The challenge of time resolved imaging amounts to sampling a spatio-temporal impulse response $I(x,y,t)$, where $x$ and $y$ are image coordinates and $t$ is the delay between emission and arrival of the light. Since the first capture of such data by Abramson in 1978 \cite{abramson1978light}, different technologies and methods for recording light-in-flight images have emerged. Streak cameras accelerate and deflect photoelectrons in order to separate them depending on their time of production. This allows very high temporal resolution but is limited to imaging one line at a time. Therefore, for two-dimensional imaging, it requires either scanning of the scene \cite{velten2013femto} (which makes data acquisition time-costly and rules out single-shot imaging) or further modification of the set-up, like adding a digital micromirror device (DMD) in order to encode the signal spatially~\cite{gao2014single,liang2017single}. To-date streak cameras provide the optimum temporal resolution with commercially available systems claiming 
resolutions of $\sim$100\,fs, but are also the most expensive of the available technologies.

Intensified charge-coupled devices (ICCD) provide high pixel counts and have recently been shown to be able to reach down to 10\,ps temporal resolution for suitably prepared scenes \cite{cester2019time}. This is, however, limited by various restrictions on the type of measured data, and requires bulky and costly hardware. 

A cheap alternative is the use of photonic mixer devices (PMD) \cite{heide2013low}, which are based on intensity modulated illumination and a special sensor pixel design that allows measuring the phase shift between outgoing and incoming illumination. They are generally used as ToF sensors for depth imaging and provide high spatial, but low temporal resolution. 

Arrays of single-photon avalanche diodes (SPAD) are rapidly becoming a leading technology for high temporal resolution imaging. This is due to the ability to manufacture time-correlated single-photon counting (TCSPC) electronics for each individual pixel directly onto the sensor chip allowing for timing resolutions on the order of tens of picoseconds \cite{Richardson2009,Henderson2018}. Currently SPAD arrays suffer from a relatively low pixel count and thus by themselves cannot be employed for many of the above imaging applications. \red{Morimoto et al. recently demonstrated a ground-breaking 1 Megapixel SPAD array with timing capabilities shown in a LIDAR type experiment \cite{Morimoto2020}. The sensor lacked TCSPC electronics however, instead gaining temporal information by scanning an electronic gate and identifying the rising edge of the response, this approach is often unsuitable for high-resolution FLIM due to the low photon efficiency.}

\begin{figure}[t]
\centering
\includegraphics[width = \columnwidth]{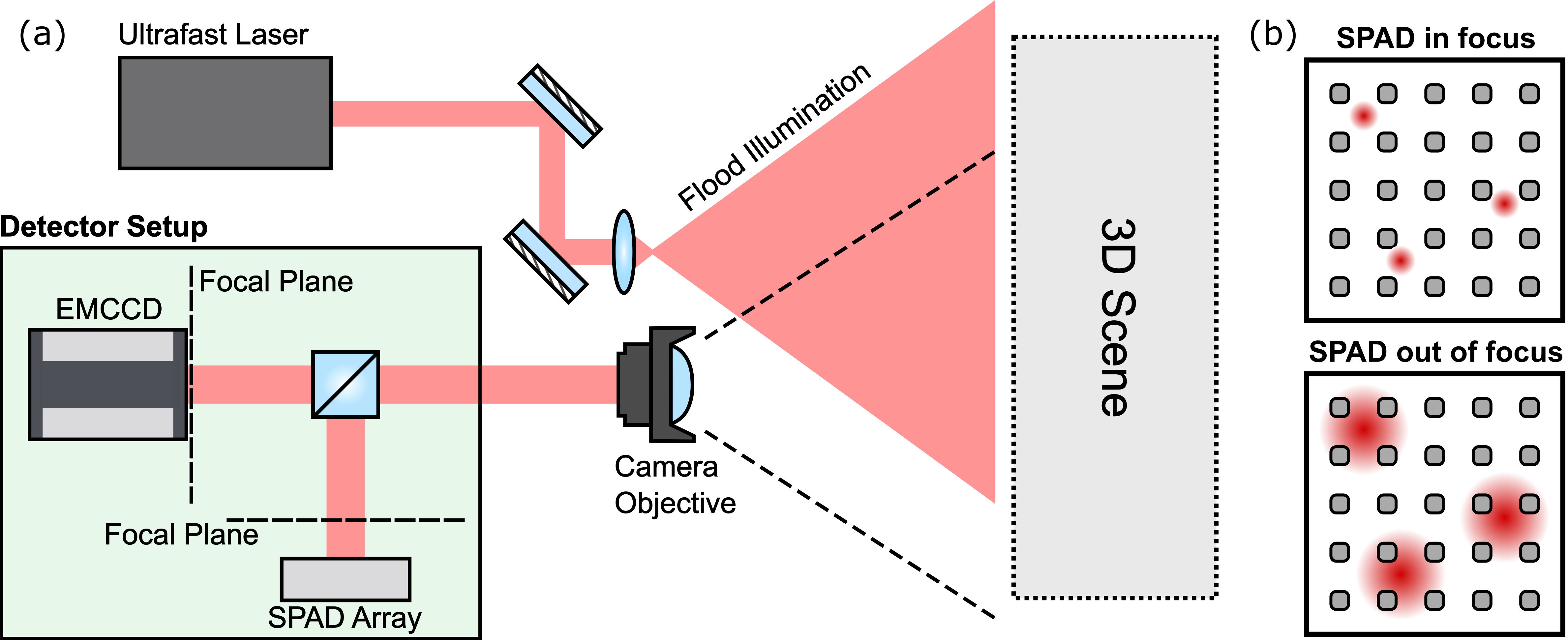}
\caption{(a) Schematic sketch of the imaging set-up for the LIDAR experiment: The scene is uniformly illuminated by pulsed laser from the same direction as the camera setup. Light is collected and then imaged onto both the high-resolution CCD array and the low-resolution SPAD array by the same objective lens. The SPAD sensor is placed slightly out of the focal plane to ensure the temporal information from each point within the scene spreads across multiple pixels. (b) Effect of the optical blur: With the low fill-factor sensor in focus such that the imaging system’s point-spread-function PSF (coloured circles) is smaller than the pixel pitch, regions of the scene are not collected by the pixels (grey areas). Shifting the sensor out of the focal plane blurs the PSF and ensures collection of the temporal information from each point in the scene.}
\label{fig:setup}
\end{figure}

In this paper, we present a method to provide three-dimensional images with both high spatial and temporal resolution. In our approach, we fuse the time-resolved image of a SPAD detector, which is low in spatial resolution, with the image of a conventional CCD sensor which integrates the signal over the whole acquisition time but provides higher spatial resolution. Our method uses an optical blur to ensure that temporal information from the entire field of view is captured by the SPAD detector despite the low fill-factor \red{typical of SPAD arrays with in-built TCSPC electronics}. In this respect the approach is similar to other methods that exploit blur for increased dynamic range, image restoration methods, or to otherwise avoid a loss of critical information \cite{Rouf2011,favaro20073}. Point spread functions optimized by artificial neural networks have also been proposed, using the neural network for the image reconstruction and the SPAD data alone \cite{Sun2020}. \red{The combination of sensor fusion and optical blur negates the heavy under-sampling of the scene caused by the pixel geometry of the SPAD array at the expense of sacrificing a percentage of the collected light intensity for the higher resolution sensor (up to 50\% in the work shown here).} By using convex optimization, a data cube with the temporal resolution of the SPAD detector and the spatial resolution of the CCD camera is reconstructed. Furthermore, our approach is capable of compensating sensor flaws like dead pixels in the SPAD array. We first verify our method with numerical simulations and assess its performance --- details of this can be found in the Supplementary Information. We then demonstrate the method practically on two different temporal imaging schemes: namely multipath LIDAR and FLIM.\\

Our method upsamples the whole three-dimensional light-in-flight image. { Similar to the works of O'Toole \cite{otoole2017reconstructing} and Lindell \cite{lindell2018towards},} our optimization acts on the data cube as a whole, not on a reduction to a two-dimensional depth map. This feature is key for applications where simple interpolation methods will yield an incorrect result, such as micron-scale changes in the florescence lifetime arising from the structure of single cells in FLIM. To test the robustness of our approach, we also demonstrate its potential using a separate publicly accessible dataset acquired with a similar experimental configuration \cite{lindell2018single} (see Supplementary Information). Our method retains high quality image reconstructions even in the presence of ambient light.\\ 

\begin{figure*}[!h]
\centering
\newcommand{\resf}{100pt}
\newcommand{\resfs}{48pt}
\begin{tabular}{cccl}
(a) & (b) & (c) & \hspace{17pt}(d) \\[3pt]
\multirow{2}{*}[40pt]{\includegraphics[height = \resf]{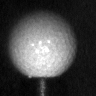}} &
\multirow{2}{*}[40pt]{\includegraphics[height = \resf]{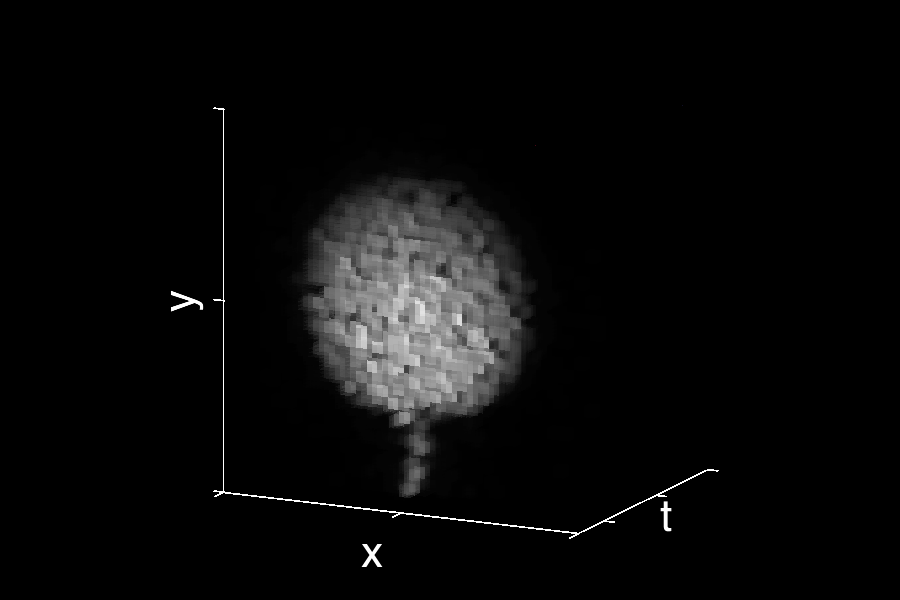}} &
\multirow{2}{*}[40pt]{\includegraphics[height = \resf]{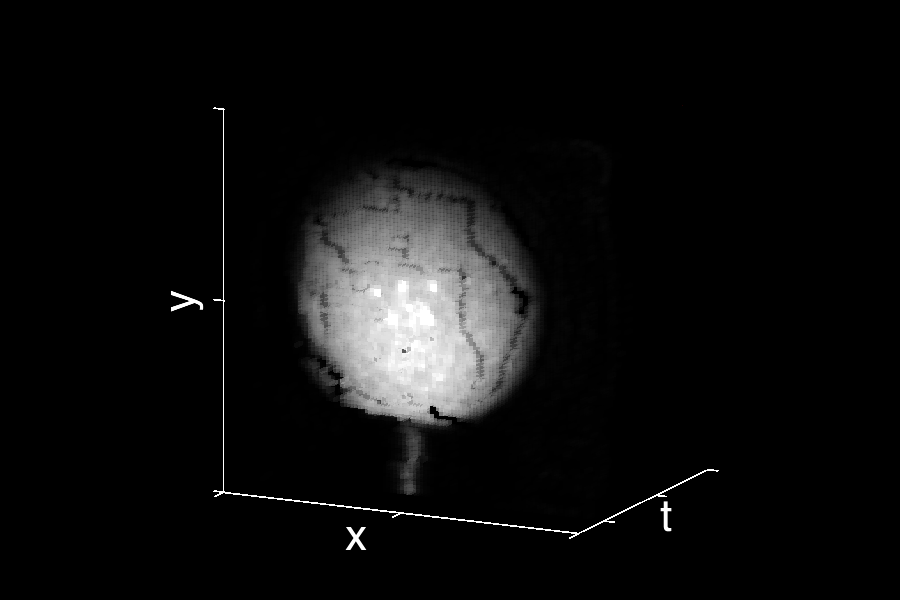}} &
\includegraphics[height = \resfs]{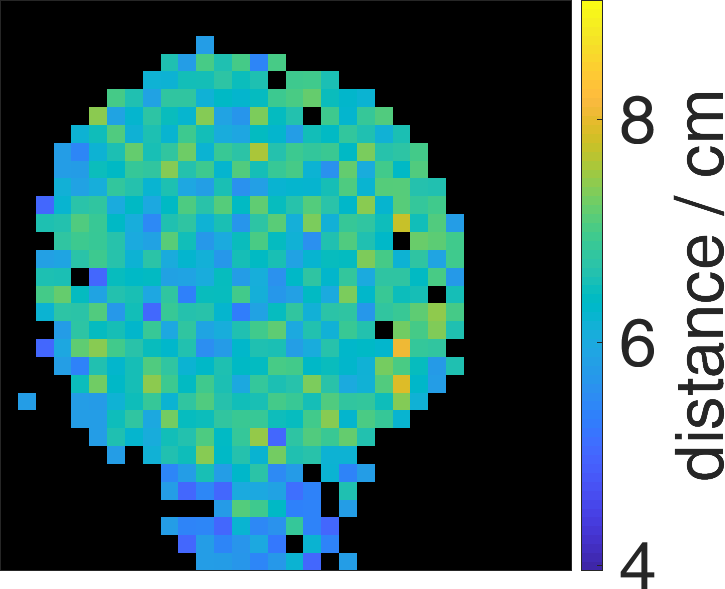} \\
& & & \includegraphics[height = \resfs]{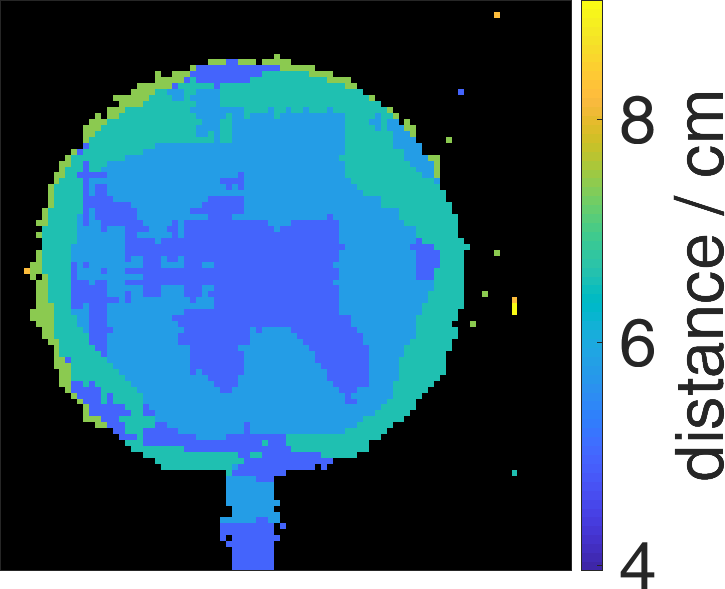} \\[2pt]

\multirow{2}{*}[40pt]{\includegraphics[height = \resf]{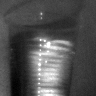}} &
\multirow{2}{*}[40pt]{\includegraphics[height = \resf]{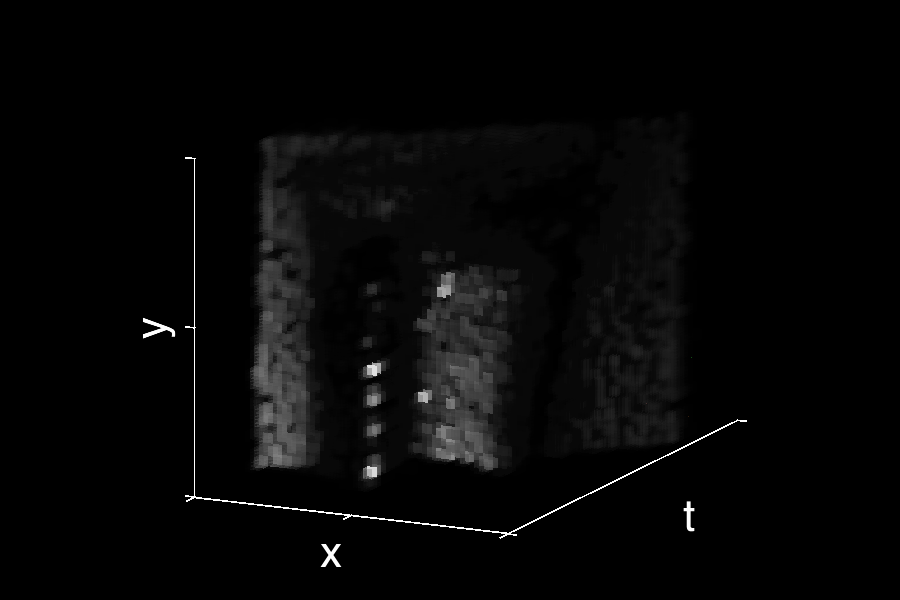}} &
\multirow{2}{*}[40pt]{\includegraphics[height = \resf]{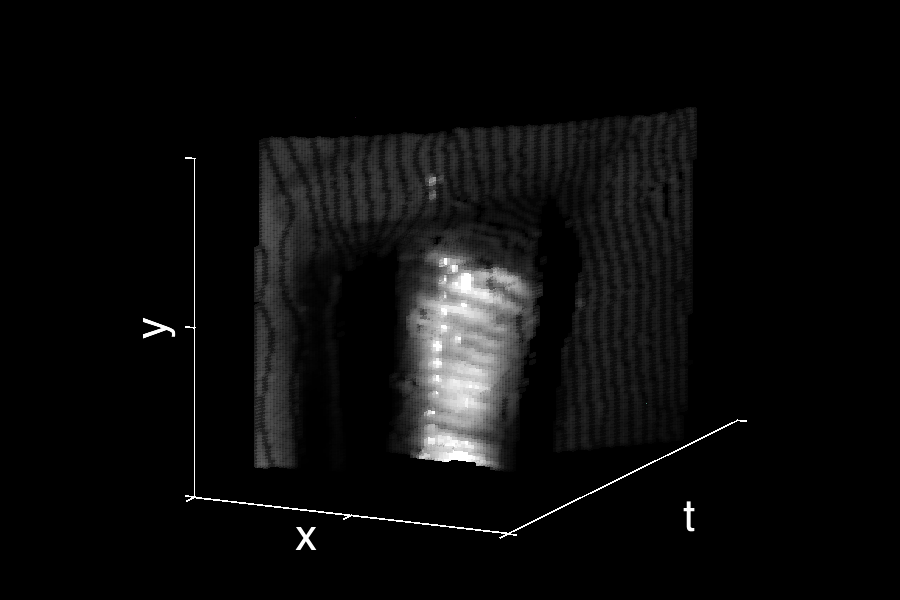}} &
\includegraphics[height = \resfs]{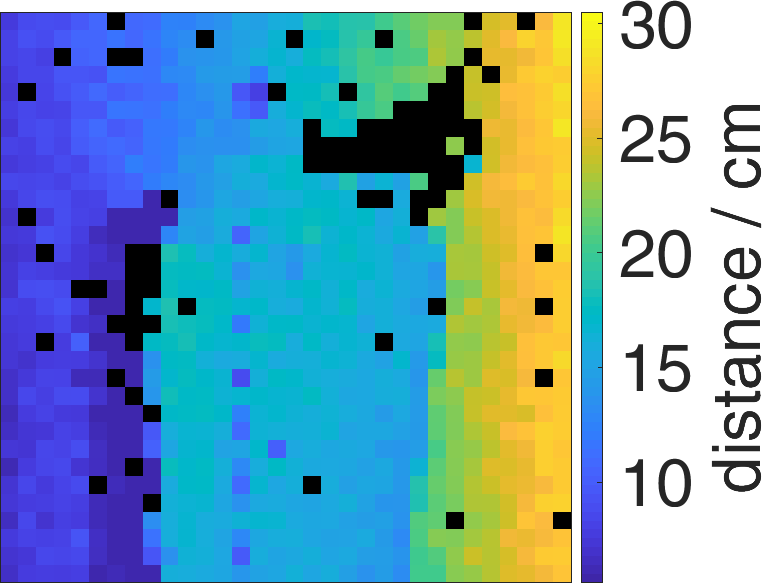} \\
& & & \includegraphics[height = \resfs]{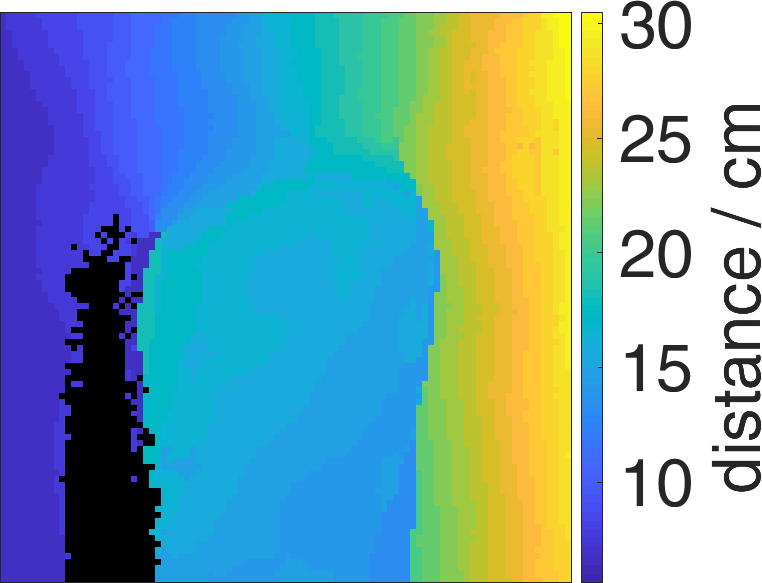} \\[2pt]

\multirow{2}{*}[40pt]{\includegraphics[height = \resf]{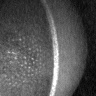}} &	
\multirow{2}{*}[40pt]{\includegraphics[height = \resf]{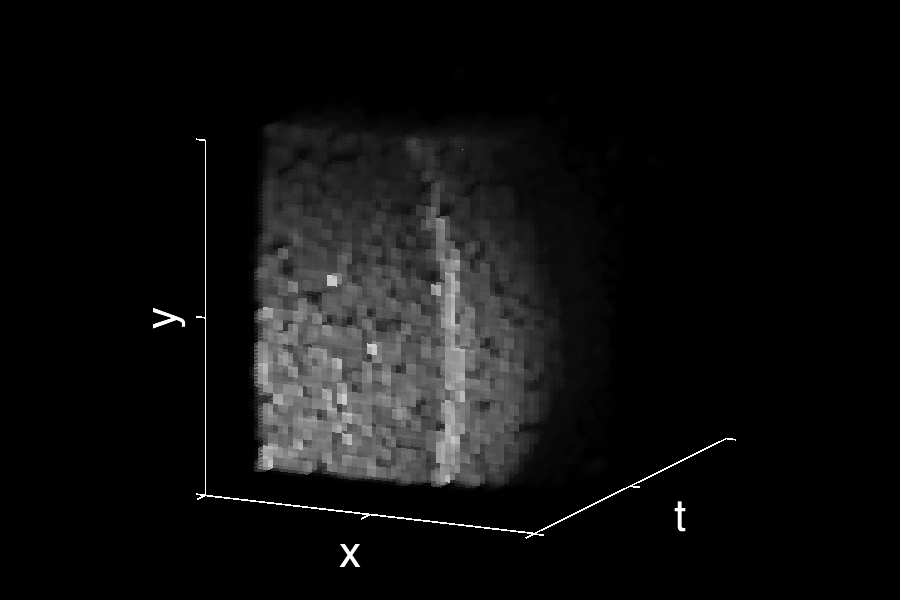}} &
\multirow{2}{*}[40pt]{\includegraphics[height = \resf]{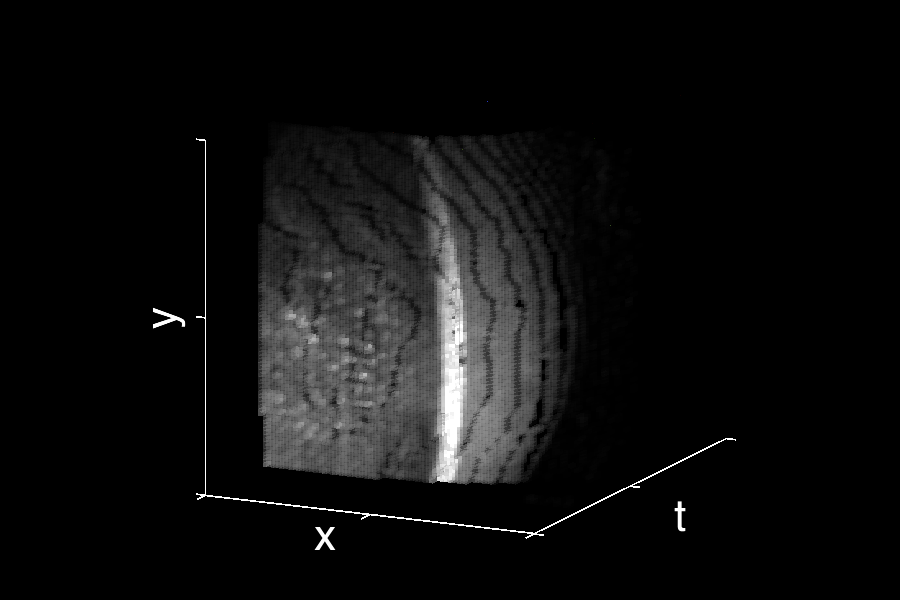}} &
\includegraphics[height = \resfs]{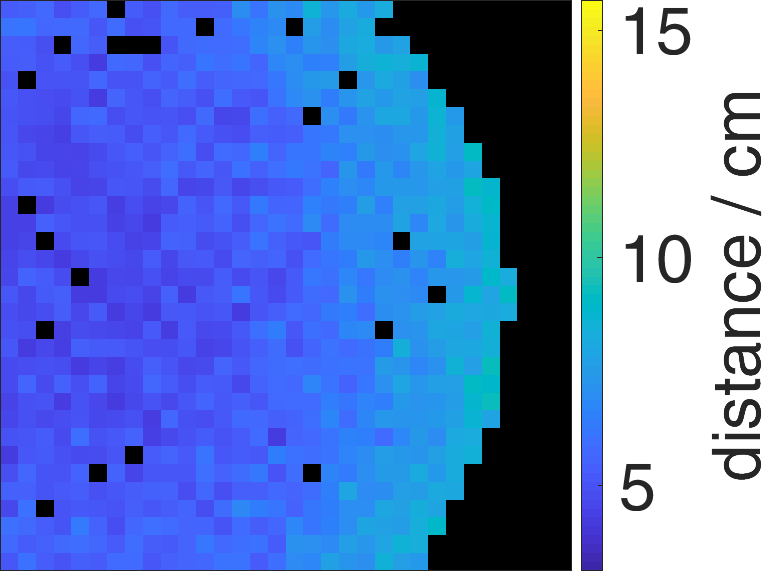} \\
& & & \includegraphics[height = \resfs]{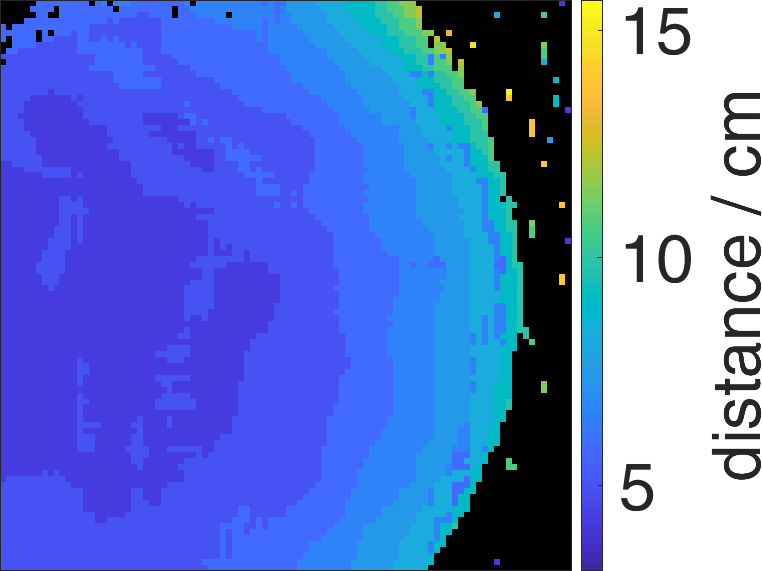}\\[2pt]

\multirow{2}{*}[40pt]{\includegraphics[height = \resf]{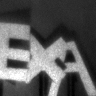}} &
\multirow{2}{*}[40pt]{\includegraphics[height = \resf]{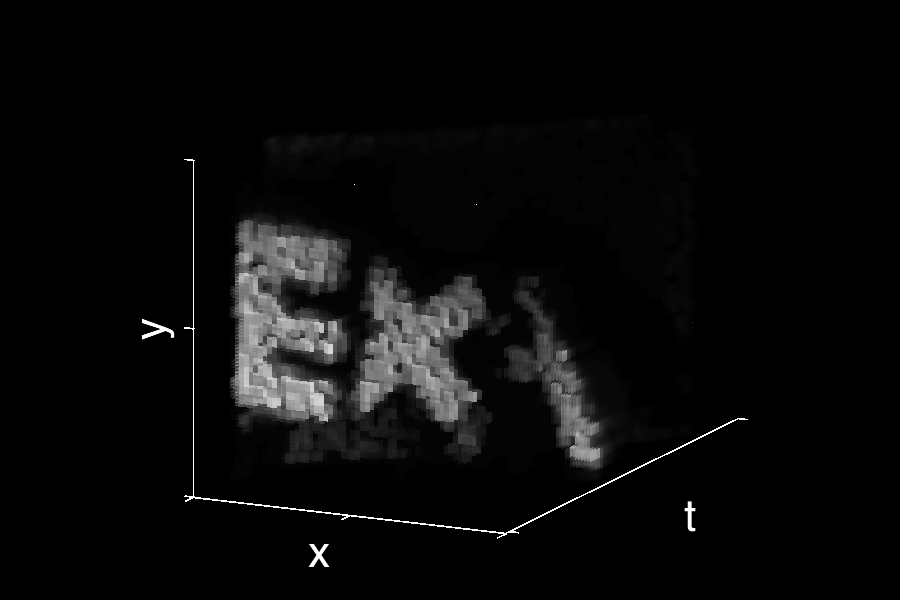}} &
\multirow{2}{*}[40pt]{\includegraphics[height = \resf]{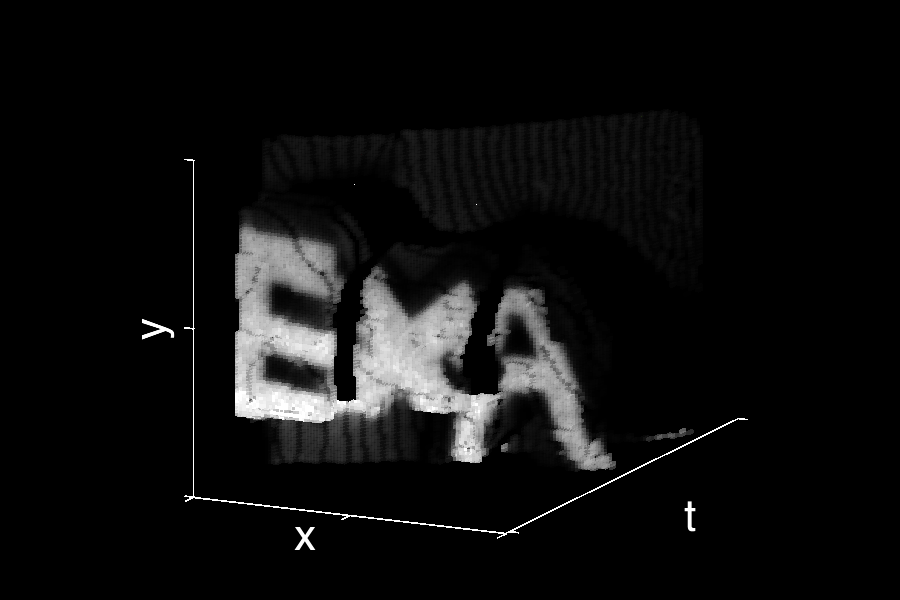}} &
\includegraphics[height = \resfs]{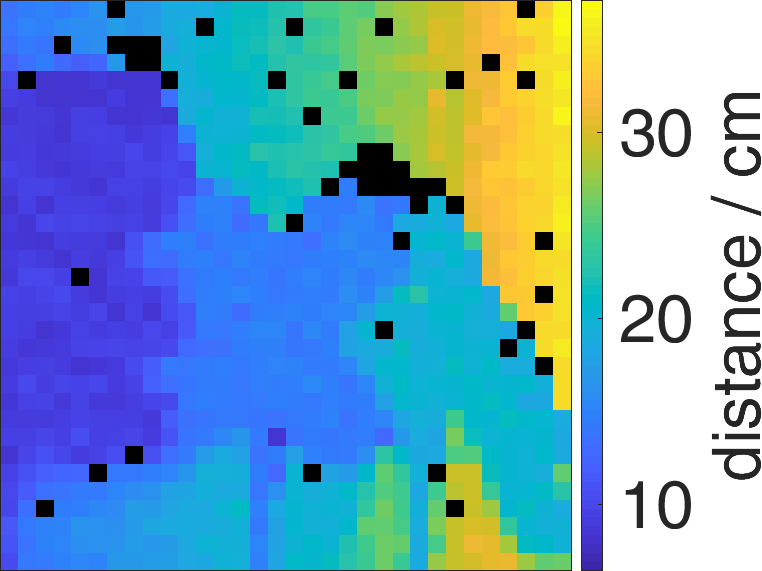} \\
& & & \includegraphics[height = \resfs]{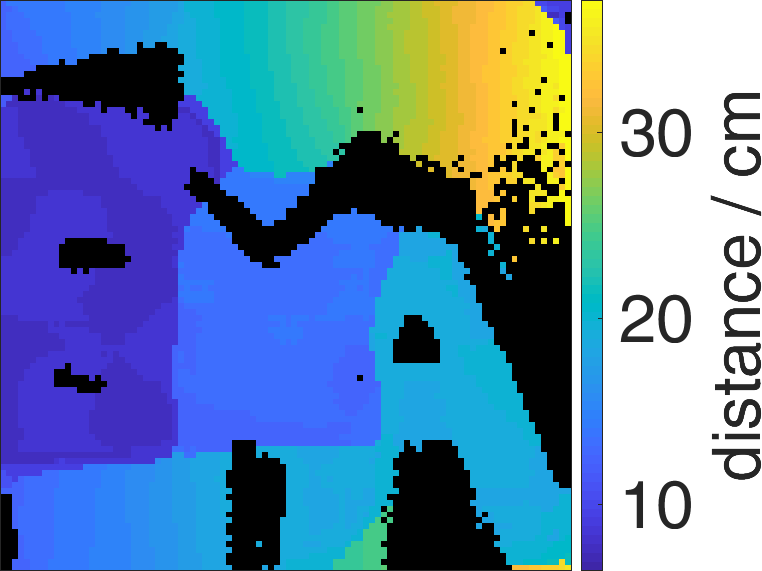} \\[2pt]

\multirow{2}{*}[40pt]{\includegraphics[height = \resf]{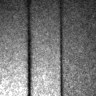}} &
\multirow{2}{*}[40pt]{\includegraphics[height = \resf]{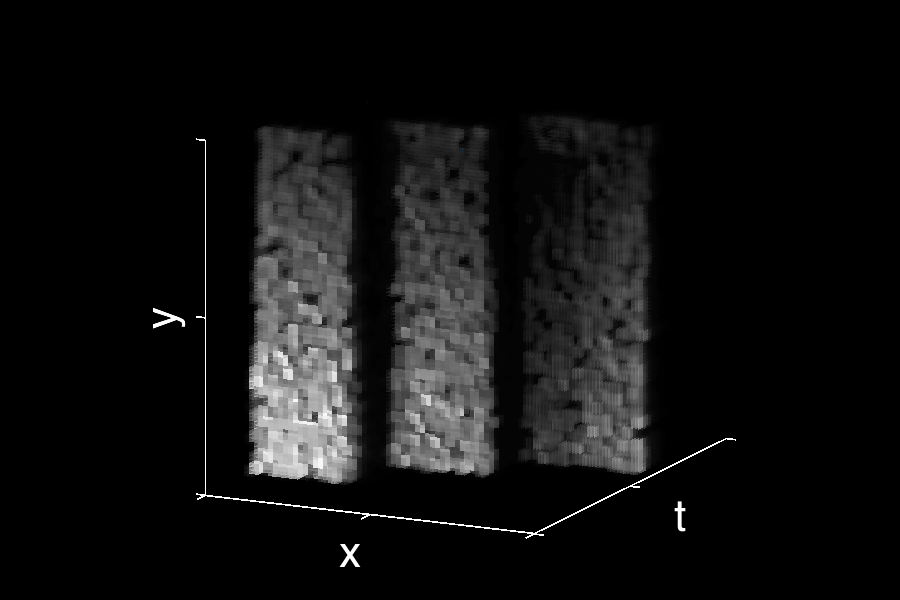}} &
\multirow{2}{*}[40pt]{\includegraphics[height = \resf]{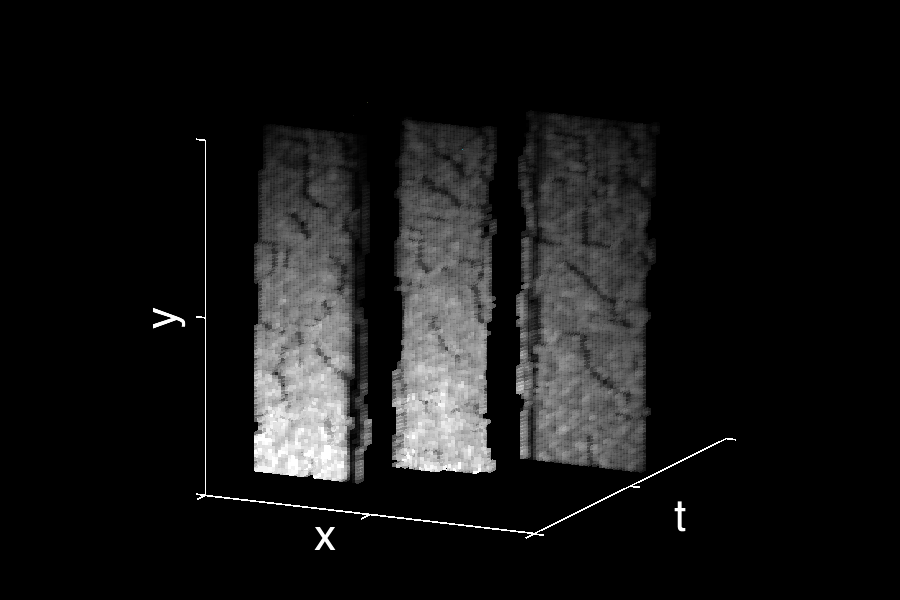}} &
\includegraphics[height = \resfs]{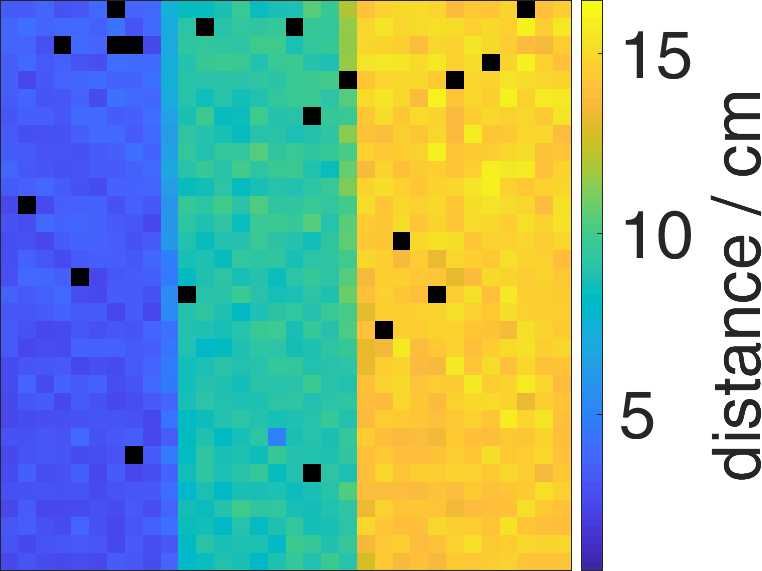} \\
& & & \includegraphics[height = \resfs]{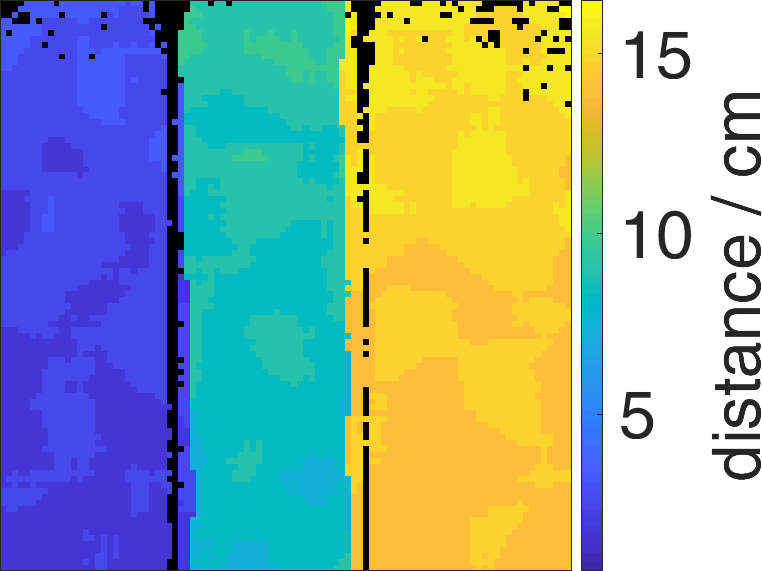} \\[2pt]
\end{tabular}
\caption{(a) CCD image, (b) $32\times 32 \times t$ SPAD measurement, (c) $96\times 96\times t$ reconstructed light-in-flight image and (d) depth images extracted from the SPAD measurement (top) and the reconstruction (bottom) for different scenes. (a) \& (b) are used as the inputs for our algorithm. From top to bottom the measurements show a golfball, a plastic cup filled with water in front of a slanted wall, detail of a basketball, three cardboard letters with a few centimetres distance between them in front of a slanted wall, three cardboard steps. Black areas in the depth images correspond to pixels with very low signal-to-noise ratio that therefore contain no meaningful depth information.}
\label{fig:results}
\end{figure*}

\section*{Computational Fusion of Sensor Data}

The goal of our method is to construct a final dataset with the spatial resolution of a high pixel density CCD sensor and the temporal resolution of a SPAD array. For a SPAD dataset of spatial resolution $m \times n$ pixels and $\tau$ timebins, and a high pixel density dataset of $M \times N$ pixels, this results in a final datacube, $i_\text{HR}(x,y,t)$, of dimensions $M \times N \times \tau$.\\
SPAD arrays typically suffer from poor fill-factor (around 1\% for the array in this work, see Methods), this results in a loss of information from light falling outside of the active areas. The image, therefore, needs to be optically filtered to prevent aliasing. We achieve this by moving the sensor slightly out of focus, such that the light from each point in the scene reaches at least one pixel's active area and we therefore capture the temporal information from each point within the scene. The resulting blur is then accounted for during the data analysis (see Supplementary Information) such that the algorithm retrieves the full $i_\text{HR}(x,y,t)$ with the correct temporal information at each spatial coordinate.\\
The forward model is designed to encapsulate these features, we represent this with a matrix:
\begin{equation}
A = P \cdot S \cdot B
\end{equation}
where $B$ performs a blurring operation to account for the defocusing, $S$ is a mask accounting for the sparse sampling of the SPAD array, and $P$ performs a spatial downsizing of the higher $M \times N$ dimensions to the lower $m \times n$ dimensions (full details in Supplementary Information). The final SPAD camera measurement, the low spatial-resolution time-of-flight image $d$, is then given by:
\begin{equation}
d = A_\tau \cdot i_\text{HR}
\end{equation}
with $A_\tau$ applying $A$ to all time bins and both $i_\text{HR}$ and $d$ being in vector form. The high-resolution transient image, $i_\text{HR}$, is reconstructed via
\begin{eqnarray} 
i_\text{HR} &=& \underset{i\in \mathbb{R}^{MN\tau}}{\mathrm{arg\,min}} \: \left\lVert{A_\tau i - d}\right\rVert_2 + \alpha  \left\lVert{Ti - c}\right\rVert_2  \nonumber \\
&& +  \beta  \left\lVert{K_h i - K_l d} \right\rVert _2+ \gamma \left\lVert{i}\right\rVert_1  + \delta  \left\lVert{\nabla_\text{2D} i} \right\rVert_1  \nonumber \\
&&\mathrm{subject\;to} \: i \geq 0 
\label{eq:opt}
\end{eqnarray} 
where $T$ performs a temporal integration over the data cube, $c$ is the vectorized CCD image and $K_h$ and $K_l$ perform a spatial integration over the high resolution and low resolution data cube, respectively. The third term enforces a similarity between the temporal distribution of photon counts in the measured data and in the reconstruction, this proved to be an essential prior in the reconstruction. The fourth term promotes sparsity of the reconstructed data cube and, while not affecting the quality of the result significantly, it ensures stability of the reconstruction. {  The last term is a 2-dimensional total variation prior that acts on the spatial dimensions of each frame, which we found to significantly improve the results for scenes with large amounts of multiply scattered light.}

The relative weights $\alpha$, $\beta$, $\gamma$ and $\delta$ have been tuned to yield the best results (see Numerical Simulations in Supplementary). The optimization is performed using CVX 2.1~\cite{cvx,gb08} and Gurobi 7.52~\cite{gurobi}\red{/MOSEK}.\\

\section*{Experimental Results: LIDAR}

We verify our method with data from a LIDAR scene depicted in Fig.~\ref{fig:setup} and described in detail in the Methods section. The raw data is first denoised and adjusted as described in the Supplementary Information. The high spatial resolution data cube is then reconstructed according to Eq.~\ref{eq:opt}.
The parameters used for this and all other data shown in this paper and the Supplementary Information are listed in Table~\ref{tab:parameters}. To model the blur of the defocused image on the SPAD sensor, a standard deviation of 6 CCD pixel widths was used. This value was found empirically as the one yielding the best reconstruction results and its accordance with the data was verified using in- and out-of-focus data acquisitions. 

%

The reconstruction results for different scenes are shown in Fig.~\ref{fig:results}. Column (a) and (b) show the raw measurements, (c) shows the reconstructed data cubes. One can see that high-frequency textures that are well visible in the CCD image but not in the SPAD measurements have been transferred into the light-in-flight image. Surfaces are much smoother, less noisy, and sharpened in all dimensions.\footnote{Black lines visible in the images are due to temporal quantisation and rendition of the data (the temporal axis is shown with a factor 3 in order to keep aspect ratios consistent).} 

In addition to the reconstructed data cubes, simple depth images of the raw SPAD measurement and reconstructed scene are shown in column (d), where the time bin with the highest photon count per spatial pixel was used as the depth value. Here, it is well visible that dead pixels from the SPAD array have been filled in, even though they have not been masked or otherwise specifically addressed before or during the reconstruction. This is possible because due to the blur, information from the dead regions is not lost, but spread over and mixed into surrounding pixels and can therefore be reconstructed. Noise has also been reduced in comparison to the raw SPAD data. \red{In general, the method remains robust to noise levels expected in a realistic experiment. A full discussion of this is presented in the Supplementary Material.}

The raw data of our SPAD measurements, as well as the reconstructed high resolution light-in-flight images rendered as videos showing the light propagating through the scene, can be found in the Supplementary Information, along with run times for all datasets.

\section*{Experimental Results: FLIM}
\begin{figure*}
\centering
\includegraphics[width = 15cm]{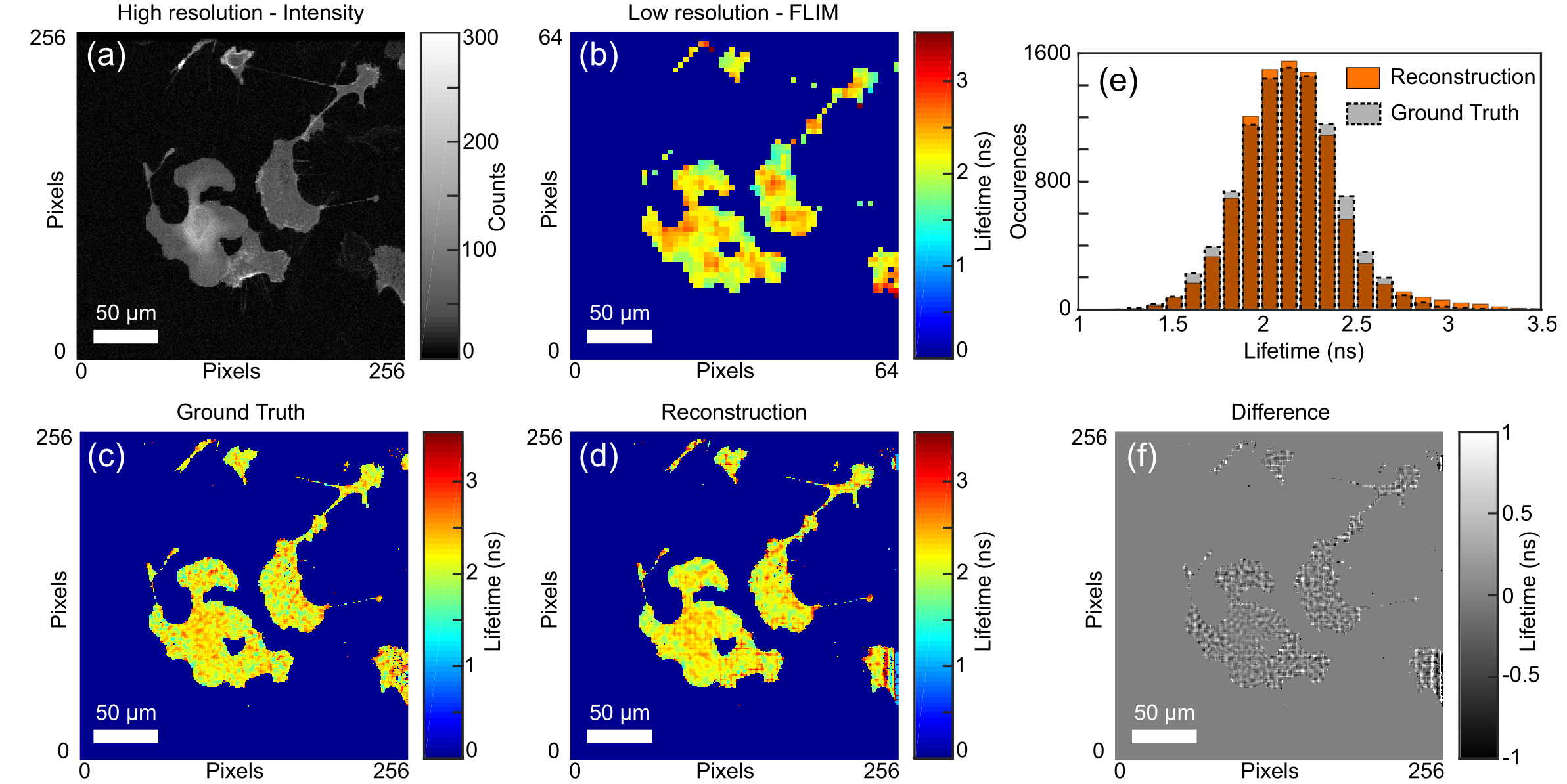}
\caption{Spatial upsampling of the fluorescence lifetime of cancer cells. (a) Full 256$\times$256 resolution intensity image. (b) Down-sampled low resolution fluorescence lifetime image. (c) Ground truth image at the full 256$\times$256 resolution. (d) Result of our reconstruction algorithm with a 4$\times$4 pixel upsampling. (e) Distribution of the reconstructed lifetimes. \red{(f) The pixel-by-pixel difference in lifetime between the Reconstruction and the Ground Truth Data.}}
\label{fig:flim_results}
\end{figure*}

\begin{figure*}
\centering
\includegraphics[width = 15cm]{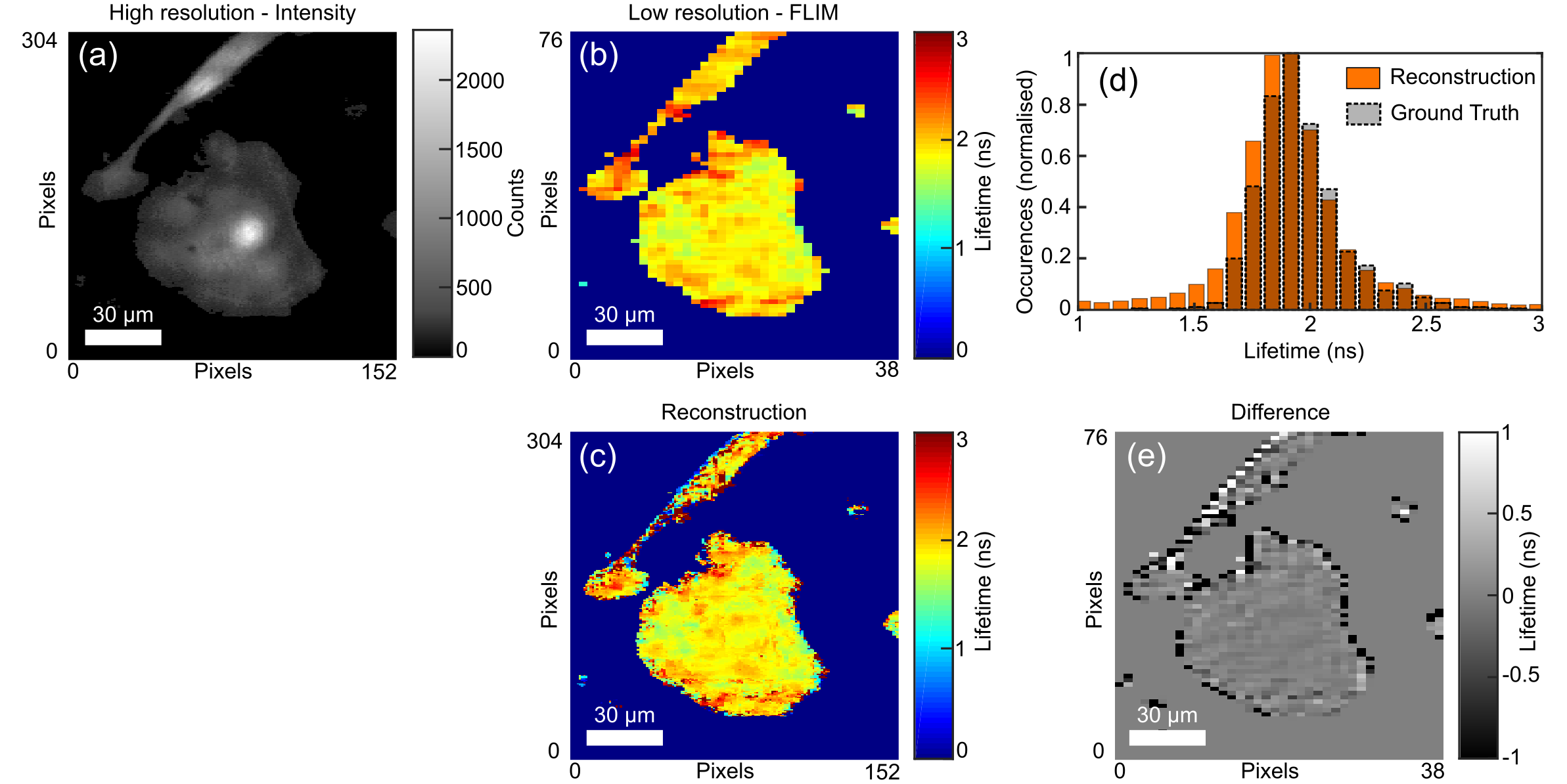}
\caption{\red{Fluorescence lifetime images of SKOV3 cancer cells expressing pcDNA3.1-mClover from a SPAD-CCD based microscope. (a) Intensity image from the CCD sensor at a spatial resolution of 304$\times$152 pixels. (b) Lifetime image obtained from the SPAD sensor with a resolution of 76$\times$38 pixels. (c) Reconstructed image upscaled by a factor of 4$\times$4 to the same resolution as (a). (d) The lifetime distributions of the reconstructed image and the low spatial resolution dataset from the SPAD sensors (Ground Truth).  (e) The pixel-by-pixel difference between low-resolution data and the reconstruction downsampled to the same pixel resolution.}}
\label{fig:flim_results_2}
\end{figure*}

We next show the potential of our method for FLIM using a commercial microscope, the details of which are described in the Methods section. The sample consists of ovarian cancer cells expressing Raichu-Rac clover-mCherry \cite{Itoh2002,Martin2018} and images are acquired using a single point scanning approach in a 256$\times$256 grid.  The temporal information is acquired with TCSPC in 75 timebins of 160 ps duration. From this data we build a lower resolution dataset that emulates the measurement that would be performed by the SPAD array. We apply our forward model operator, $A_\tau$, following Eq.~2 with a downsampling ratio of 4$\times$4. This results in a 64$\times$64$\times$75 temporally resolved dataset that forms the low-resolution input to our algorithm, $d$. For the high spatial resolution dataset, $c$, we take the total time-integrated photon counts from the full 256$\times$256 pixel array to form an intensity image. The lifetimes are estimated by fitting a single exponential decay model to both $d$ and $i_\text{HR}$, bounded between 1 ns and 7 ns using prior knowledge of the lifetime distribution. The algorithm input images, along with the resulting lifetime image, are depicted in Fig.~\ref{fig:flim_results}. We test the validity of our approach by comparing the distribution of the measured lifetimes with those obtained with the algorithm, shown in Fig.~\ref{fig:flim_results}e . There is a high level of fidelity to the ground truth data with the overall shape of the ground truth distribution. \red{Artefacts in the reconstructed image can be observed in the lower right corner, these are most likely due to blurring operator, $B$, incorporating pixels that lay outside of the 256$\times$256 pixel field-of-view, thus including areas which were physically measured. We note that this situation does not exist in when an optical blur is included in the setup, as opposed to the synthetic one introduced here.} The algorithm mean lifetime of the reconstructed image was 2.18 ns with a standard deviation 0.41 ns, in close agreement with the ground truth lifetimes of (2.14 $\pm$ 0.25) ns. The same approach could also be used to improve the acquisition speed of point scanning imaging systems such as the one used to acquire the data in Fig.~\ref{fig:flim_results}. Our results show a reduction in the number of time-resolved measurements (spatial points) by at least a factor of 16 can be achieved with the amount of time needed to acquire the high resolution image being small in comparison. We note, however, that substantial time is still required for the reconstruction post-measurement. \red{Whilst the temporal information within a LIDAR scene can often be readily approximated with a linear interpolation, this is rarely the case for more complex systems such as FLIM where there may be non-trivial sub-micron changes in the lifetime. Our method can, however, account for these small scale variations by acquiring temporal information from the whole scene and by interpreting using the high-resolution spatial information as an additional constraint.}\\
\red{We next demonstrate the potential of our approach on a bespoke SPAD camera-based FLIM microscope, the details of which can be found in the methods section. Here, unlike the previous examples, the field-of-view of the SPAD sensor is matched onto the CCD (Andor something) with a $\times$5.55 smaller magnification thereby allowing for a higher pixel density whilst also demonstrating the robustness of the approach when different optical systems are used for the two sensors. We note that in this case, ground-truth lifetime data cannot be obtained in the same way as shown in Fig~.\ref{fig:flim_results}. Fig.~\ref{fig:flim_results_2} shows images taken of cancer cells (details in Methods) and the corresponding reconstruction for a 4$\times$4 upsampling from the SPAD array. There are clearly features in the lifetime image which cannot be fully resolved by the lower pixel count SPAD sensor but become apparent in the high resolution reconstruction such as the small size and distinct shape of the longer lifetime regions at e.g. in the center of the larger cell. For comparison, the distribution of the lifetimes for the low-resolution SPAD data and the final reconstruction are shown, displaying a strong level of similarity. The pixel-wise difference is also calculated by binning 4$\times$4 pixel areas of the reconstructed dataset before evaluating the lifetime to match the dimensions of the SPAD data. This is displayed in Fig.~\ref{fig:flim_results_2}e and shows good agreement between the two datasets with artefacts arising at the cell boundaries due to the 4$\times$4 binning.}

\section*{Conclusion}

Our method shows that with a simple optical set-up and a conventional camera, the spatial resolution of a SPAD array sensor can be increased significantly. In simulations, a factor of 12 could be achieved on each spatial dimension, corresponding to a factor of 144 in pixel count, even in the presence of noise. Low fill factor limitations could be overcome by moving the SPAD sensor slightly out of focus. \red{In this way, the proposed method has the potential to push the spatial resolution of time-resolved SPAD arrays well beyond the current state of the art and into the few mega pixel domain.} Holes in the SPAD measurement due to dead pixels are filled in by the reconstruction. This has been demonstrated on measurement data with an upsampling factor of 3$\times$3 on LIDAR data and a factor of 4$\times$4 on FLIM data due to hardware limitations. On additional datasets that have not been captured with our hardware set-up we demonstrated upsampling of $8 \times 8$ and $16 \times 16 $ after blurring and downsampling the original $256 \times 256$ pixels SPAD data. These results, as well as those from simulated measurements, suggest that using a CCD sensor with higher pixel density, our method would allow higher upsampling factors also with our original hardware set-up. Additionally, our method is not limited by a low signal-to-noise ratio or the presence of ambient light as evidenced by the reconstructions in the Supplementary Information (``Upsampling Results on Other Data Sets" Section).

The main limitation of our method is the long run time of the reconstructions, which scales with the size of the reconstruction as well as the original SPAD measurement. { A full calibration of the light transport matrix, which would include all optical effects for the specific hardware set-up accurately, might yield even better results on experimental data. On the other hand, it would supposedly also make the model more bound to a specific set-up, and less flexible in the application to new unknown hardware systems. However, it could be a worthwhile enhancement for a fixed (commercial) system.} Considering the availability of small form factor SPAD and CCD sensors, both could be combined into a single, convenient device.

\red{\section*{Data Accessibility}
All data and code used in this article can be found at \url{https://github.com/ccallenberg/spad-ccd-fusion}}

\section*{Methods}
\subsection*{Experimental Setup: LIDAR}
For the high temporal resolution dataset, we use a 32$\times$32 SPAD array with in-pixel Time Correlated Single Photon Counting (TCSPC) capabilities of 55\,ps bin width \cite{gariepy2015single,Richardson2009}. The sensor layout consists of 7\,$\mu$m diameter sensors with a 50\,$\mu$m pitch and is of the same basic design now commercialised by Photon Force Ltd. Exposure times of up to 13\,s are used. The high spatial resolution is obtained using an Andor iXon emCCD with a 512$\times$512 pixel array cropped to 96$\times$96 pixels to match the field of view of the SPAD array. \red{The same optics are used to match the SPAD and emCCD field of views ensure that the two cameras can be co-registered with minimal aberrations/defects. This achieved using a 50:50 beamsplitter so that all of the collect light is directed to one of the two sensors. CO-registering in this way does not however present a limitation of out approach as illustrated by Fig.~\ref{fig:flim_results_2} which was acquired with different optical components and magnification for each sensor.} The emCCD is used without gain such that it operates as a conventional CCD. Exposure times of the order of 100\,ms are used. The same camera objective (12\,cm fisheye) is used for both sensors in parallel, separated with a beamsplitter. The illumination source is Ti:Sapph oscillator of 130\,fs pulse duration at a repetition rate of 80\,MHz and a centre wavelength of 800\,nm which flood illuminates the scene. The SPAD camera acquisition is synchronised with the laser using an Optical Constant Fraction Discriminator to minimise electronic jitter.\\
\subsection*{Multiphoton time-domain fluorescence lifetime imaging (FLIM)}
The following process was used to prepare the data shown in Fig.~\ref{fig:flim_results}. Cells were left to equilibrate on a heated microscope insert at 37\,$^\circ$C, perfused with 5\,\% CO$_2$ prior to imaging.  Images were acquired in the dark using a multiphoton LaVision TRIM scan head mounted on a Nikon Eclipse inverted microscope with a 20X water objective.  Illumination is provided by a Ti:Sapphire femtosecond laser used at 920\,nm (12\,\% power).  Clover signal was passed through band pass filters 525/50\,nm emission and acquired using a \red{FLIM X-16 Bioimaging Detector TCSPC FLIM system (LaVision BioTec)}.  A 254\,$\mu$m$^2$ field of view correlating to 256 pixel$^2$ was imaged at 600 Hz with a 10 line average \red{in a total acquisition time of 5199 ms.}\\
\red{For the data shown in Fig.~\ref{fig:flim_results_2}, a bespoke microscope system was created using and Andor Zyla as the high-spatial resolution CMOS array and the 192$\times$128 pixel FLIMera system developed by HORIBA Scientific based upon the chip presented in \cite{Henderson2018}. The microscope consisted of a $\times$20 0.4 NA Olympus objective with a 250\,mm focal length tube lens for the Zyla and a 50\,mm tube lens for the SPAD sensor, resulting in a $\times$5.5 magnification at the SPAD sensor compared to the Zyla. Before passing the data into the reconstruction algorithm the images from the Zyla were downsampled to a size of 608$\times$304 pixels from which a 304$\times$152 pixel area was taken to match a 76$\times$38 pixel area selected from the SPAD field of view. Exposure time was 500\,ms.}

\subsection*{Mammalian cell lines, culturing conditions and transfections}
SKOV3 cells were maintained in Dulbecco’s modified Eagle’s medium (DMEM) supplemented with 10\,\% FBS, 2\,mM L-Glutamine and 1X PenStrep.  Cell lines were maintained in 10 cm dishes at 37\,$^\circ$C and 5\,\% CO$_2$.
SKOV3 cells were transfected in the morning using Amaxa Nucleofector (Lonza) kit V, program V-001 with \red{either} 5 $\mu$g Raichu-Rac1\_Clover-mCherry \red{or pcDNA3.1-mClover} DNA (adapted from \cite{Itoh2002}) following manufacturers guidelines and replated on 6 cm TC-treated dishes at 37\,$^\circ$C and 5\,\% CO$_2$.
\red{For live cell imaging,} cells were collected and replated onto 35\,mm glass bottom MatTek dishes that were previously coated overnight with laminin (10\,$\mu$g\,ml$^{-1}$) diluted in PBS.  These were left overnight at 37\,$^\circ$C, 5\,\% CO$_2$.  
The next morning prior to imaging, the dishes were washed twice with pre-warmed PBS and replaced with pre-warmed FluoroBrite DMEM supplemented with 10\,\% FBS, 2\,mM L-Glutamine and 1X PenStrep.
\red{For fixed cell imaging, the cells were collected and replated onto 22 mm glass coverslips that were previously coated overnight with laminin (10 $\mu\text{gml}^{-1}$) diluted in PBS.  These were left overnight at 37 $^\circ$C, 5\% CO$_2$.  The next day, these cells were fixed in 4\% PFA for 10 minutes and washed with PBS and mounted using Fluromount-G (Southern Biotech).}

\subsection*{Reconstruction parameters}

\begin{table}[h!]
\caption{Parameters used in the reconstruction of all images in the main text and Supplementary Information.}
\begin{ruledtabular}
\begin{tabular}{lllll}
 & $\alpha$ & $\beta$ & $\gamma$ & $\delta$\\\colrule 
\rule{0pt}{3ex}LIDAR (Fig.\,2) & \red{$1$} & \red{$10^{-4}$} & \red{$10^{-2}$} & \red{$0$}\\
FLIM (Fig.\,3) 				   & $1$ & $10^{-3}$ & $10^{-7}$ & $10^{-5}$\\
Sim. monkey (Suppl. Fig.\,6-7) & \red{$1$} & \red{$10^{-7}$} & \red{$10^{-4}$} & \red{$0$}\\
Sim. table (Suppl. Fig.\,9-10) & \red{$1$} & \red{$10^{-7}$} & \red{$10^{-4}$} & \red{$10^{-7}$}\\
\end{tabular}
\end{ruledtabular}

\label{tab:parameters}
\end{table}

\subsection*{Acknowledgements}

The authors acknowledge funding from EPSRC (UK, grant no. EP/T00097X/1 and EP/T002123/1) and the European Research Council (ERC Starting Grant "ECHO"). DF is supported by the Royal Academy of Engineering under the Chairs in Emerging Technologies scheme.

\bibliographystyle{unsrt}
\bibliography{references}

\begin{thebibliography}{10}

\bibitem{gariepy2015single}
Genevieve Gariepy, Nikola Krstaji{\'c}, Robert Henderson, Chunyong Li, Robert~R
  Thomson, Gerald~S Buller, Barmak Heshmat, Ramesh Raskar, Jonathan Leach, and
  Daniele Faccio.
\newblock Single-photon sensitive light-in-fight imaging.
\newblock {\em Nature communications}, 6:6021, 2015.

\bibitem{Velten2012}
Andreas Velten, Thomas Willwacher, Otkrist Gupta, Ashok Veeraraghavan,
  Moungi~G. Bawendi, and Ramesh Raskar.
\newblock {Recovering three-dimensional shape around a corner using ultrafast
  time-of-flight imaging}.
\newblock {\em Nature Communications}, 3:745, 2012.

\bibitem{Buttafava2015}
Mauro Buttafava, Jessica Zeman, Alberto Tosi, Kevin Eliceiri, and Andreas
  Velten.
\newblock {Non-line-of-sight imaging using a time-gated single photon avalanche
  diode}.
\newblock {\em Optics Express}, 23(16):20997, 2015.

\bibitem{Faccio2018}
Daniele Faccio and Andreas Velten.
\newblock {A trillion frames per second: The techniques and applications of
  light-in-flight photography}.
\newblock {\em Reports on Progress in Physics}, 81(10), 2018.

\bibitem{Durduran2010}
T.~Durduran, R.~Choe, W.~B. Baker, and A.~G. Yodh.
\newblock {Diffuse optics for tissue monitoring and tomography}.
\newblock {\em Reports on Progress in Physics}, 73(7), 2010.

\bibitem{Satat2016}
Guy Satat, Barmak Heshmat, Dan Raviv, and Ramesh Raskar.
\newblock {All Photons Imaging Through Volumetric Scattering}.
\newblock {\em Scientific Reports}, 6(September):1--8, 2016.

\bibitem{Lyons2019b}
Ashley Lyons, Francesco Tonolini, Alessandro Boccolini, Audrey Repetti, Robert
  Henderson, Yves Wiaux, and Daniele Faccio.
\newblock {Computational time-of-flight diffuse optical tomography}.
\newblock {\em Nature Photonics}, 13(8):575--579, aug 2019.

\bibitem{Dowling1999}
K~Dowling, M~J Dayel, S~C~W Hyde, P~M~W French, M~J Lever, J~D Hares, and A~K~L
  Dymoke-Bradshaw.
\newblock {High resolution time-domain fluorescence lifetime imaging for
  biomedical applications}.
\newblock {\em Journal of Modern Optics}, 46(2):199--209, 1999.

\bibitem{abramson1978light}
Nils Abramson.
\newblock Light-in-flight recording by holography.
\newblock {\em Optics letters}, 3(4):121--123, 1978.

\bibitem{velten2013femto}
Andreas Velten, Di~Wu, Adrian Jarabo, Belen Masia, Christopher Barsi, Chinmaya
  Joshi, Everett Lawson, Moungi Bawendi, Diego Gutierrez, and Ramesh Raskar.
\newblock Femto-photography: capturing and visualizing the propagation of
  light.
\newblock {\em ACM Transactions on Graphics (ToG)}, 32(4):44, 2013.

\bibitem{gao2014single}
Liang Gao, Jinyang Liang, Chiye Li, and Lihong~V Wang.
\newblock Single-shot compressed ultrafast photography at one hundred billion
  frames per second.
\newblock {\em Nature}, 516(7529):74, 2014.

\bibitem{liang2017single}
Jinyang Liang, Cheng Ma, Liren Zhu, Yujia Chen, Liang Gao, and Lihong~V Wang.
\newblock Single-shot real-time video recording of a photonic mach cone induced
  by a scattered light pulse.
\newblock {\em Science advances}, 3(1):e1601814, 2017.

\bibitem{cester2019time}
Lucrezia Cester, Ashley Lyons, Maria Braidotti, and Daniele Faccio.
\newblock Time-of-flight imaging at 10 ps resolution with an iccd camera.
\newblock {\em Sensors}, 19(1):180, 2019.

\bibitem{heide2013low}
Felix Heide, Matthias~B Hullin, James Gregson, and Wolfgang Heidrich.
\newblock Low-budget transient imaging using photonic mixer devices.
\newblock {\em ACM Transactions on Graphics (ToG)}, 32(4):45, 2013.

\bibitem{Richardson2009}
Justin~A. Richardson, Lindsay~A. Grant, and Robert~K. Henderson.
\newblock {Low dark count single-photon avalanche diode structure compatible
  with standard nanometer scale CMOS technology}.
\newblock {\em IEEE Photonics Technology Letters}, 21(14):1020--1022, 2009.

\bibitem{Henderson2018}
Robert~K. Henderson, Nick Johnston, Haochang Chen, David Day~Uei Li, Graham
  Hungerford, Richard Hirsch, David McLoskey, Philip Yip, and David~J.S. Birch.
\newblock {A 192$\times$128 Time Correlated Single Photon Counting Imager in
  40nm CMOS Technology}.
\newblock {\em ESSCIRC 2018 - IEEE 44th European Solid State Circuits
  Conference}, pages 54--57, 2018.

\bibitem{Morimoto2020}
Kazuhiro Morimoto, Andrei Ardelean, Ming-Lo Wu, Arin~Can Ulku, Ivan~Michel
  Antolovic, Claudio Bruschini, and Edoardo Charbon.
\newblock {Megapixel time-gated SPAD image sensor for 2D and 3D imaging
  applications}.
\newblock {\em Optica}, 7(4):346, apr 2020.

\bibitem{Rouf2011}
Mushfiqur Rouf, Rafal Mantiuk, Wolfgang Heidrich, Matthew Trentacoste, and
  Cheryl Lau.
\newblock {Glare encoding of high dynamic range images}.
\newblock In {\em CVPR 2011}, pages 289--296. IEEE, jun 2011.

\bibitem{favaro20073}
P~Favaro and S~Soatto.
\newblock {\em {3-D Shape Estimation and Image Restoration}}.
\newblock Springer London, London, 2007.

\bibitem{Sun2020}
Qilin Sun, Jian Zhang, Xiong Dun, Bernard Ghanem, Yifan Peng, and Wolfgang
  Heidrich.
\newblock {End-to-end Learned, Optically Coded Super-resolution SPAD Camera}.
\newblock {\em ACM Transactions on Graphics}, 39(2):1--14, apr 2020.

\bibitem{otoole2017reconstructing}
Matthew O'Toole, Felix Heide, David~B Lindell, Kai Zang, Steven Diamond, and
  Gordon Wetzstein.
\newblock Reconstructing transient images from single-photon sensors.
\newblock In {\em Proceedings of the IEEE Conference on Computer Vision and
  Pattern Recognition}, pages 1539--1547, 2017.

\bibitem{lindell2018towards}
David~B Lindell, Matthew O'Toole, and Gordon Wetzstein.
\newblock Towards transient imaging at interactive rates with single-photon
  detectors.
\newblock In {\em 2018 IEEE International Conference on Computational
  Photography (ICCP)}, pages 1--8. IEEE, 2018.

\bibitem{lindell2018single}
David~B Lindell, Matthew O'Toole, and Gordon Wetzstein.
\newblock Single-photon 3d imaging with deep sensor fusion.
\newblock {\em ACM Transactions on Graphics (TOG)}, 37(4):113, 2018.

\bibitem{cvx}
Michael Grant and Stephen Boyd.
\newblock {CVX}: Matlab software for disciplined convex programming, version
  2.1.
\newblock \url{http://cvxr.com/cvx}, March 2014.

\bibitem{gb08}
Michael Grant and Stephen Boyd.
\newblock Graph implementations for nonsmooth convex programs.
\newblock In V.~Blondel, S.~Boyd, and H.~Kimura, editors, {\em Recent Advances
  in Learning and Control}, Lecture Notes in Control and Information Sciences,
  pages 95--110. Springer-Verlag Limited, 2008.

\bibitem{gurobi}
{Gurobi Optimization, LLC}.
\newblock Gurobi optimizer reference manual, 2018.

\bibitem{Itoh2002}
Reina~E. Itoh, Kazuo Kurokawa, Yusuke Ohba, Hisayoshi Yoshizaki, Naoki
  Mochizuki, and Michiyuki Matsuda.
\newblock {Activation of Rac and Cdc42 Video Imaged by Fluorescent Resonance
  Energy Transfer-Based Single-Molecule Probes in the Membrane of Living
  Cells}.
\newblock {\em Molecular and Cellular Biology}, 22(18):6582--6591, sep 2002.

\bibitem{Martin2018}
Kirsty~J. Martin, Ewan~J. McGhee, Juliana~P. Schwarz, Martin Drysdale,
  Saskia~M. Brachmann, Volker Stucke, Owen~J. Sansom, and Kurt~I. Anderson.
\newblock {Accepting from the best donor; analysis of long-lifetime donor
  fluorescent protein pairings to optimise dynamic FLIM-based FRET
  experiments}.
\newblock {\em PLoS ONE}, 13(1):1--25, 2018.

\end{thebibliography}

\end{document}